\documentclass{article}
\usepackage{graphicx}

\usepackage{geometry}
\usepackage{float}
\usepackage[backend=biber,sorting=none]{biblatex}
\usepackage{gensymb}
\usepackage{color,soul}
\usepackage[dvipsnames]{xcolor}

\usepackage[affil-it]{authblk}

\title{Nanophotonic dispersion engineering in tantala-based microresonators with ultralow-loss claddings}

\author[1]{Alexa R. Carollo \thanks{alexa.carollo@nist.gov}}
\author[1,2]{Atasi Dan}
\author[1]{Jizhao Zang}
\author[1,2]{Haixin Liu}
\author[1,2]{Nitesh Chauhan}
\author[1]{Scott B. Papp}

\affil[1]{Time and Frequency Division, National Institute of Standards and Technology, Boulder, CO 80305, USA}
\affil[2]{Department of Physics, University of Colorado Boulder, Boulder, CO 80309, USA}

\begin{document}

\maketitle

\noindent Integrated photonics optimized to leverage low loss, nonlinear-optical processes and optoelectronic integration requires robust optical cladding materials and associated fabrication process flows. We develop low-loss silicon dioxide (SiO$_2$) claddings with tantalum pentoxide (Ta$_2$O$_5$, tantala) integrated photonics to realize robust, high intrinsic quality factor ($Q_i$) microresonators and photonic-crystal resonators (PhCRs) for soliton microcomb generation. Specifically, we introduce a full-wafer, low-temperature process compatible with two tantala-based films: pure tantala and an amorphous metal-oxide mixture of titanium dioxide (TiO$_2$, titania) and tantala. Combining tantala with titania enhances the film's resistance to damage from the SiO$_2$ cladding deposition process, offering access to high $Q_i$ above $4 \times 10^6$ in fully cladded microresonators. Our platform supports low-loss edge couplers, group-velocity dispersion engineering by waveguide geometry, access to nanophotonic structures particularly to control phase-matching, and high dimensional tolerance in devices across an entire wafer. Using the platform, we explore the generation of high-efficiency, dark-soliton microcombs with repetition frequency from 50 GHz to 500 GHz, which supports applications such as photonic artificial intelligence acceleration, electronic signaling, and optical data communication.

\section{Introduction}

Kerr nonlinear optics in microresonators offers powerful controls and access to physical behaviors with light, enabling explorations of novel phenomena, development of signaling and information processing, and generation of light sources for applications from artificial intelligence (AI) to quantum science. Supercontinuum generation in integrated waveguides provides access to broadband frequency combs \cite{Carlson:19} for optical frequency metrology \cite{Carlson:PRA2017}, spectroscopy \cite{Nader:2018, Baumann:19}, and imaging \cite{Ji:2021}. Kerr microresonators enable optical parametric oscillation \cite{PhysRevLett.93.083904}, four-wave mixing frequency translation \cite{Black:2023}, and generation of discrete, driven--dissipative soliton excitations that provide frequency comb sources directly in integrated photonics \cite{spencer_optical-frequency_2018,drake_terahertz-rate_2019}. Soliton microcombs, in particular, have been explored for photonic AI acceleration \cite{xu_11_2021}, nonclassical light generation \cite{Guidry:23}, and data communication \cite{pirmoradi_integrated_2025, shirpurkar_photonic_2022}.

The use of Kerr nonlinear optics generally requires access to low-loss waveguide or resonator devices to reach the threshold for parametric oscillation and to provide phase matching. Kerr integrated nonlinear photonics that support substantial refractive index contrast for the design of phase matching has been explored in a range of materials. Among these materials, thick films of stoichiometric silicon nitride obtained via low-pressure chemical vapor deposition are a reference, offering exceptionally low optical loss and high microresonator intrinsic quality factor, $Q_i$, above $1\times10^7$, even when fabricated via a complementary metal-oxide-semiconductor (CMOS) process on 300 mm diameter wafers \cite{Liu:25}. However, processing such silicon nitride films requires mitigation of high tensile stress and high temperature annealing at 1200 \degree C \cite{Liu2021}, thus motivating the development of alternative materials with more favorable fabrication requirements. Tantalum pentoxide (Ta$_2$O$_5$, tantala) has unique advantages over silicon nitride, including lower temperature processing, low tensile stress, and higher Kerr nonlinearity, which open up a range of capabilities in integrated photonics \cite{Jung:21}. For example, tantala films are deposited at room temperature via ion-beam sputtering (IBS), enabling direct deposition onto thermally-sensitive substrates like thin-film lithium niobate for visible photonics applications \cite{Brodnik2026}. Furthermore, the ability to tailor the composition of tantala films is a key advantage. In an amorphous metal oxide mixture of titanium dioxide (TiO$_2$, titania) and tantala, the oxygen vacancy defect density is suppressed relative to pure tantala, which was originally found to be instrumental for reducing loss in mirror coatings for LIGO \cite{Pinard:17}. For integrated photonics applications, the improved properties of the titania-tantala mixture enable microresonators with ultra-high $Q_i$ up to $1\times10^7$ for low-threshold power optical parametric oscillation and supercontinuum \cite{Carollo:2026}.

To leverage the properties of the tantala photonics platform towards practical applications, it is important to use a low-loss protective cladding on top of the waveguide layer. Silicon dioxide (SiO$_2$) is a suitable cladding material due to its large transparency window, low optical loss, and particularly, its relatively low index of refraction ($n \approx 1.44$ at 1550 nm) \cite{Malitson:65}. The cladding layer should have a low refractive index relative to tantala, to ensure that sufficient index contrast is maintained for dispersion engineering in integrated devices. Furthermore, SiO$_2$ is compatible with standard CMOS fabrication processes, making it a promising material for scalable photonic integration. Indeed, SiO$_2$ is ubiquitous in integrated photonics as a low-loss cladding, enabling low-confinement microresonators \cite{Puckett2021}, multi-layer integration \cite{Girardi:23, MacFarlane:2019}, packaging \cite{Lamee:20}, high-yield wafer-scale fabrication \cite{Liu2021}, thermo-optic tuning \cite{Lopez:2025}, and unique dispersion engineering opportunities in slot waveguides \cite{Zhang:13} and nanocomposites \cite{Liu:2025}. Dense, high-quality SiO$_2$ films are typically grown via thermal oxidation, but this process requires furnace temperatures near 1000 \degree C \cite{Lee2012} that are incompatible with tantala. These temperatures far exceed tantala's crystallization temperature of $\approx 700$ \degree C, above which the film begins to transition from an amorphous to crystalline phase \cite{Fazio:20}. While plasma-enhanced chemical vapor deposition (PECVD) offers a viable low-temperature alternative \cite{Qiu:2023}, the process must be carefully optimized to achieve conformal, void-free coatings without introducing optical loss or mechanical stress that could compromise device integrity.

\begin{figure*}[!t]
\centering
    \includegraphics[scale=0.34]{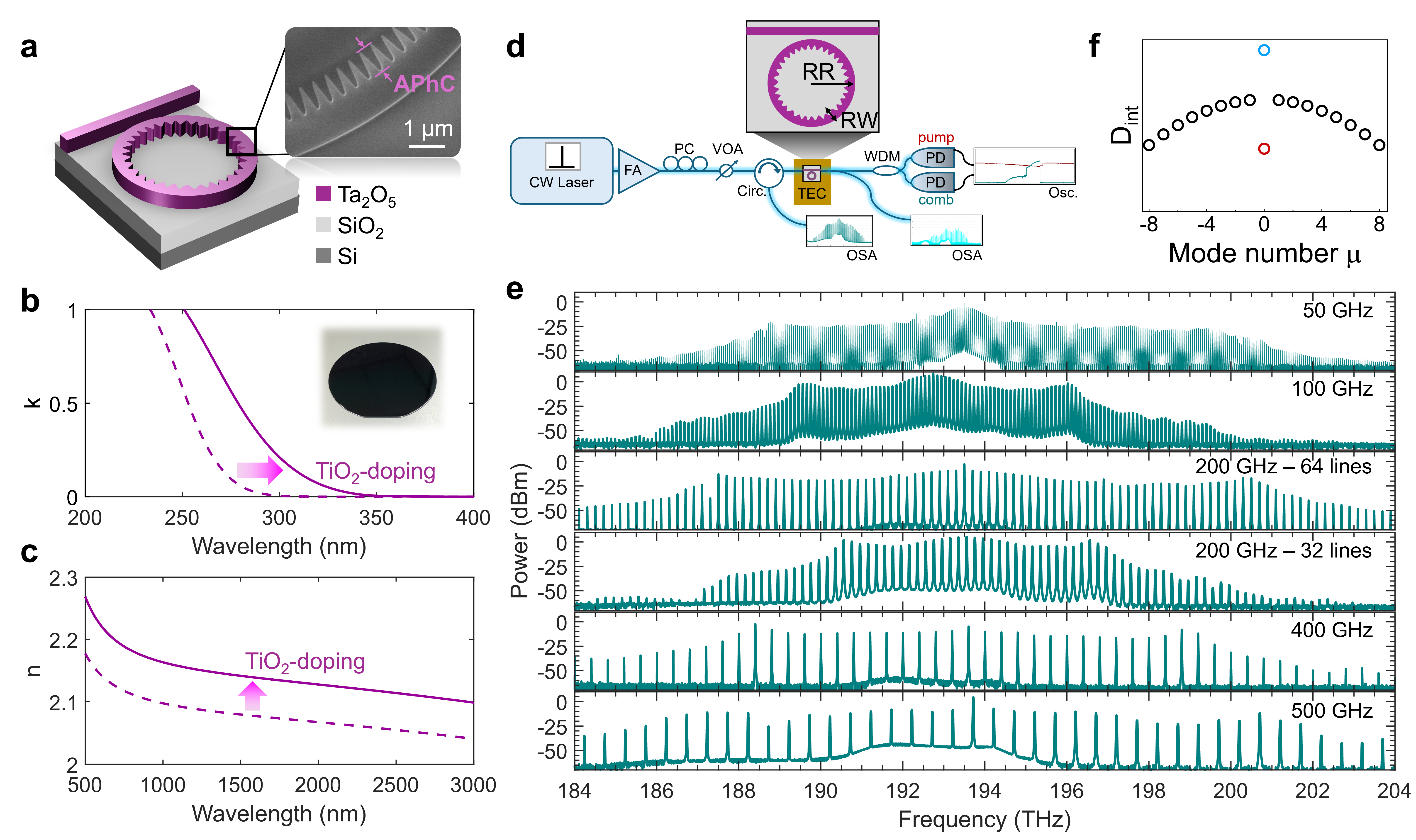}
    \caption{Overview of the tantala nanophotonics platform. (a) Illustration of a PhCR, and SEM image of a $RW$ modulation that induces a photonic bandgap. The modulation amplitude, $APhC$, is indicated in the SEM. The film layer structure is pure tantala (Ta$_2$O$_5$), silicon dioxide (SiO$_2$), and silicon (Si). (b) Extinction coefficient, $k$, and (c) index of refraction, $n$, as a function of wavelength for pure tantala (dashed line) and titania-tantala (solid line) films. Inset is a photograph of a wafer coated with titania-tantala. (d) Experimental setup for generating soliton microcombs using an amplified continuous wave (CW) laser. FA: fiber amplifier, PC: polarization controller, VOA: variable optical attenuator, Circ: circulator, TEC: thermoelectric cooler, WDM: wavelength division multiplexer, PD: photodiode, OSA: optical spectrum analyzer, Osc.: oscilloscope. (e) Microcomb spectra generated when pumping a series of tantala PhCRs with different geometries to control the repetition rate and spectral bandwidth. (f) Illustration of the effect of the PhC on $D_{\rm{int}}$. }
    \label{Fig1}
\end{figure*}

Here, we explore control over a range of soliton microcomb dynamical properties in fully SiO$_2$-clad tantala-based microresonators. Tantala-based refers to either pure tantala or amorphous metal-oxide mixtures of tantala and titania, and we use our ability to adjust the film composition as a way to optimize device performance. To explore the flexibility of soliton microcomb formation in tantala photonic-crystal resonators (PhCRs) \cite{jin_bandgap-detuned_2025}, we systematically chart the accessible parameter space in repetition rate and spectral bandwidth. With that baseline of device performance, we develop a versatile nanoscale fabrication process with low-temperature SiO$_2$ cladding deposition via inductively-coupled plasma PECVD (ICP-PECVD), ensuring compatibility with tantala-based films. In particular, the titania-tantala mixture proves better at withstanding the SiO$_2$ plasma deposition environment than pure tantala, motivating our work to develop the dispersion engineering landscape in titania-tantala. Since we need high resolution refractive index data for design, here we present index measurements and evaluation of precise geometric dispersion engineering. In fully SiO$_2$-clad PhCRs, we generate soliton microcombs with high efficiency and low threshold powers, owing to our ultralow-loss devices. These results demonstrate the flexibility in configurations offered by the tantala platform, both in choice of cladding and choice of material composition, and its potential towards enabling robust, fully integrated microcomb-based systems.

\section{Results}

Figure 1 summarizes the material properties of tantala-based films and demonstrates the generation of normal-dispersion soliton microcombs in PhCRs over a wide range of repetition rates. An attractive property of the tantala platform is the ability to fabricate PhCRs with nanometer-scale precision, which we leverage throughout this work to achieve robust Kerr phase matching; see Fig. \ref{Fig1}a. In these devices, modulation of the ring width ($RW$) induces a photonic bandgap, enabling highly designable azimuthal mode structure for phase matching in soliton microcomb generation and other Kerr nonlinear optical processes. We first consider the material properties of pure tantala. The platform exhibits a wide transparency window extending into the ultraviolet (UV) wavelength range, indicated by its low extinction coefficient, $k$, at wavelengths above 300 nm; see Fig. \ref{Fig1}b. With IBS at room temperature performed by Plasma Optik, we deposit thick, crack-free, amorphous films on thermally oxidized silicon wafers. The IBS process enables the deposition of both pure tantala and amorphous mixtures of tantala and titania, providing a route to tune optical and material properties. Titania-tantala films, also referred to as titania-doped tantala, offer several advantages over pure tantala. For integrated photonics applications, this includes intrinsically lower defect density, leading to higher $Q_i$ and enhanced nonlinear performance \cite{Carollo:2026}. These films also exhibit modified optical properties compared to pure tantala. The absorption onset is red-shifted from 300 nm to 350 nm, and the refractive index is increased by a few percent; see Fig. \ref{Fig1}c. Indeed, the optical properties are dependent on the amount of titania that the tantala films are doped with \cite{Fazio:20}. Because the group-velocity dispersion (GVD) of PhCRs depends on both geometric and material contributions, the change in refractive index upon doping with titania must be accounted for in microresonator design.

The combination of low loss and high nonlinearity of the tantala platform enables microcomb generation with a continuous wave pump laser; see Fig. \ref{Fig1}d. A tunable continuous wave pump laser is amplified using an erbium-doped fiber amplifier and coupled to the chip with a lensed fiber. We set the pump polarization with a fiber polarization controller and pump power with a variable optical attenuator. A fiber circulator separates the backward-propagating microcomb signal, which we record with an optical spectrum analyzer (OSA). A portion of the forward-propagating signal is also recorded with an OSA, and the remaining signal is used to analyze the comb formation dynamics. To analyze the comb formation dynamics, the signal is filtered with a wavelength division multiplexer to separate the pump power from the comb power, measured with a photodiode, and recorded with an oscilloscope as a function of laser detuning. During the measurement, the temperature of the chip is controlled with a thermoelectric cooler.

Through careful design of the geometry of PhCRs, primarily by varying the $RW$ and ring radius ($RR$) across devices, we control the spectral properties of microcombs; see Fig. \ref{Fig1}e. We demonstrate an exceptionally wide range of control over the repetition rate by way of the free-spectral range. For instance, as we increase the $RR$ from 54.4 \textmu m to 436 \textmu m, we achieve repetition rates ranging from 500 GHz down to 50 GHz. For a given $RR$, we control the spectral bandwidth, or the number of comb lines with high power, by varying the $RW$. For $RR$ near 109 \textmu m, we tune the spectrum for high power across 32 lines ($RW = 4$ \textmu m) or 64 lines ($RW = 2.25$ \textmu m). A smaller $RW$ supports flatter, but still normal, GVD for broader combs. Here, our designs are based on dispersion engineering to achieve normal GVD in 570 nm thick pure tantala devices with air cladding. We use refractive index data and finite-element analysis \cite{Black:21} to simulate GVD in air-clad tantala devices to target normal dispersion \cite{Spektor:2024} across a range of $RR$. While microcombs are intrinsically phase-matched in anomalous GVD microresonators, phase matching in normal dispersion microresonators requires a defect in the mode structure; see the integrated dispersion, $D_{\rm{int}}$, in Fig. \ref{Fig1}f. In an otherwise normal dispersion device, we engineer a defect at the pump mode using a photonic crystal (PhC). The PhC induces coherent backscattering to split the pump mode, with relative mode number $\mu=0$, into two modes. Pumping the lower-frequency PhC mode (red point in Fig. \ref{Fig1}f) induces a Kerr shift of the adjacent modes that is higher than that of the pump mode, resulting in phase matching of spontaneously generated microcombs \cite{Yu:2021}. These results establish tantala PhCRs as a versatile microcomb source, thus motivating our work to further develop this platform to be compatible with SiO$_2$ cladding.

\begin{figure*}[!t]
\centering
    \includegraphics[scale=0.34]{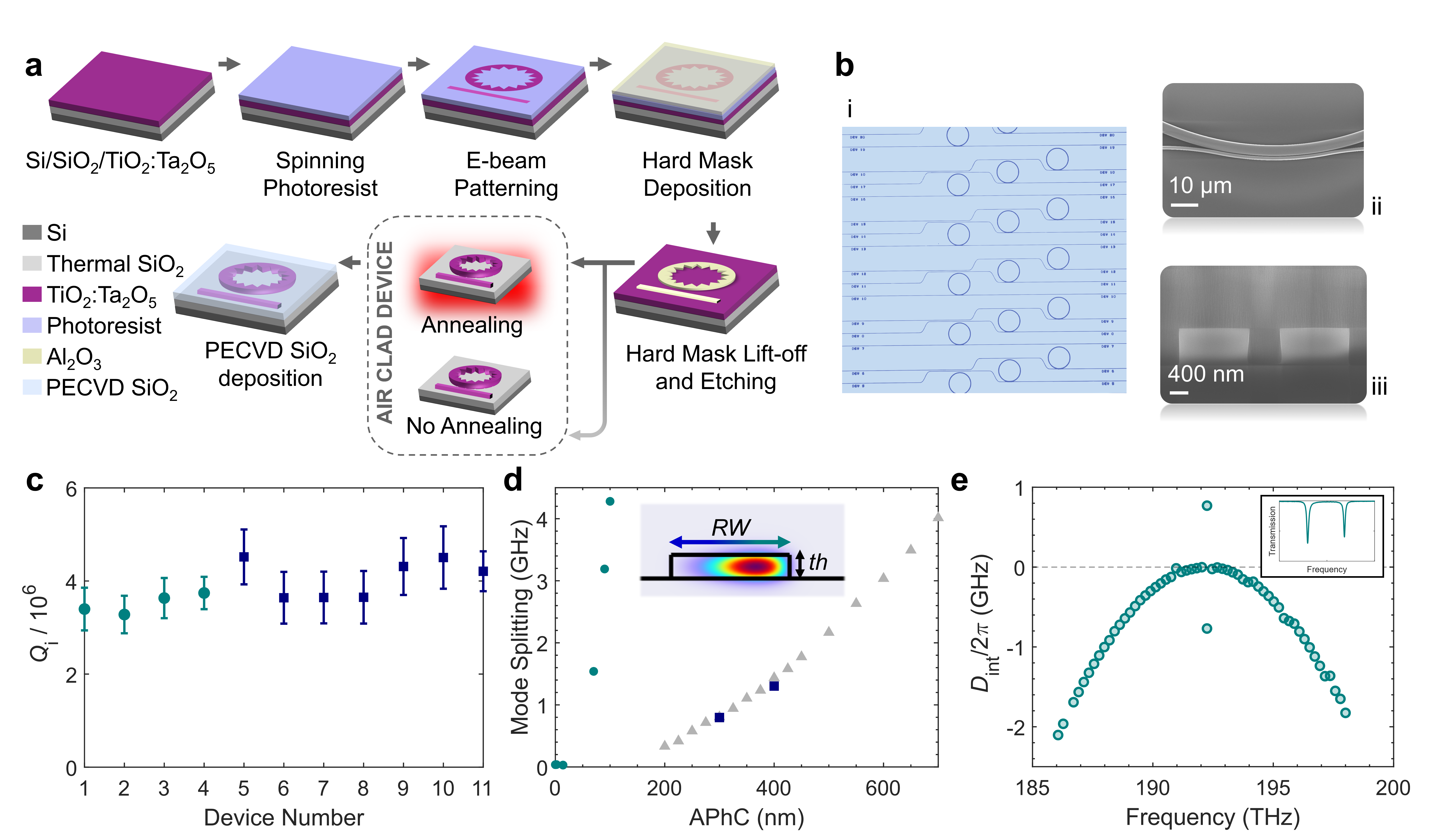}
    \caption{Fabrication and characterization of tantala-based PhCRs with ultralow-loss SiO$_2$ cladding. (a) Fabrication process flow, illustrated for titania-tantala (TiO$_2$:Ta$_2$O$_5$). (b) SiO$_2$-clad device imaging. i: Optical microscope image of a chip. ii: SEM image of a pulley coupler. iii: SEM cross section of waveguide test structures. (c) Microresonator $Q_i$ of several different titania-tantala devices. We plot average $Q_i$ of several resonance modes across the 1514 nm - 1614 nm wavelength range, and plot the standard deviation as error bars. Green circles are devices with outer $RW$ modulation, and blue squares are devices with inner $RW$ modulation. (d) Mode splitting induced by the PhC as a function of modulation amplitude, $APhC$, inscribed on the outer (green circles) or inner (blue squares) $RW$ of titania-tantala microresonators. Gray triangles show the trend for inner $RW$ modulation in a pure tantala PhCR, for comparison. Inset is the simulated electric field amplitude in a titania-tantala microresonator. (e) Measured $D_{int}$ as a function of mode frequency for a titania-tantala PhCR. Inset shows the photonic bandgap.}
    \label{Fig2}
\end{figure*}

To expand the versatility of tantala photonics towards applications requiring robust, long-term operation, we explore the addition of an SiO$_2$ top cladding layer. Our full fabrication flow with SiO$_2$ cladding is illustrated in Fig. \ref{Fig2}a. The fabrication flow begins with pure tantala films, or titania-tantala films in which a 26 \% volume fraction of titania is co-sputtered along with tantala. The film thickness is 800 nm, which is thicker than what we use for air-clad devices, to compensate for the reduced index contrast. With electron-beam lithography, we define nanopatterned designs in photoresist, and transfer the pattern into an alumina (Al$_2$O$_3$) hard mask via liftoff. After liftoff, we etch the tantala-based film into waveguides and PhCRs using fluorine-based ICP reactive ion etching (RIE). Tantala-based films are susceptible to forming oxygen vacancy defects, particularly during IBS film growth, but we recover the oxygen content by annealing the wafer at 500 \degree C for 10 hours in air after patterning. It is critical to perform this step prior to SiO$_2$ deposition. The order of these steps is important because the mechanism of reducing defect density in tantala-based films relies on oxygen diffusion from the atmosphere into the film, and thick claddings may prevent oxygen from reaching the film. We take our annealed wafer and deposit 3 \textmu m of SiO\textsubscript{2} via ICP-PECVD at 300 \degree C. Here, we use a silane (SiH$_4$) precursor, and operate at 1600 W ICP power and 300 W RF power to promote sufficient gap filling. After dicing into chips with deep RIE, a final anneal at 500 \degree C drives out material defects from the SiO$_2$.

Through the course of our work developing this procedure, we find that pure tantala is unfortunately susceptible to damage from the ICP-PECVD cladding deposition process, making it difficult to realize the same control over soliton dynamics that we have in air-clad devices. The harsh plasma environment drives out oxygen and causes an increase in defect density that cannot be fully recovered with annealing, possibly because the thick SiO$_2$ cladding layer acts as a barrier that prevents oxygen diffusion into the tantala film. This degrades $Q_i$ and affects the nonlinear performance of the film \cite{Carollo:2026}. However, we overcome this limitation with titania-tantala. Titania-tantala does not experience an increase in oxygen-vacancy defect density to the same extent as pure tantala, making it much more robust to a variety of fabrication conditions. Moreover, since it is a tantala-based film with similar material properties, the fabrication process is compatible with both films. Because of its advantages, the majority of the work described hereafter is focused on titania-tantala. We note that all of the steps prior to SiO$_2$ deposition are common to air and SiO$_2$ clad devices. With air cladding, and titania-tantala in particular, the annealing requirement is less stringent \cite{Carollo:2026}.

The fabrication process with full SiO$_2$ cladding yields numerous microresonators and integrated waveguide couplers per chip; see Fig. \ref{Fig2}b. Our integrated waveguide pulley couplers support selective coupling of the TE0 fundamental mode into the resonator. We achieve void-free filling of the SiO$_2$ into densely spaced structures, as shown in the scanning electron microscope (SEM) image of a cross section of two waveguides separated by 700 nm. Typically, we achieve void-free filling of gaps as small as 400 nm, which is important for adequately cladding the waveguides and PhCR nanostructures.

With full SiO$_2$ cladding, we achieve ultralow-loss titania-tantala microresonators with high yield; see Fig. \ref{Fig2}c. Moreover, these data presented are from devices off of two different chips, demonstrating reproducibility that is important for compatibility with wafer-scale fabrication. The additional SiO$_2$ material layer is expected to contribute some absorption in the 1550 nm wavelength region, but it remains low enough to support an average $Q_i$ up to $4.5\times10^6$. Across 11 PhCR with various $RW$ modulation amplitudes, $APhC$, on either the inner or outer ring edge, we maintain high $Q_i$. Depending on the microresonator ring structure, modulating the inner or outer wall will have a different effect on the mode splitting induced by the PhC; see Fig. \ref{Fig2}d. We explain this with a simulation of the mode profile in a microresonator with $RW=4$ \textmu m and $RR=100$ \textmu m. The mode profile of the propagating light travels close to the outer edge of the ring; see inset of Fig. \ref{Fig2}d. Thus, inscribing a modulation of the outer $RW$ is more efficient at achieving a desired mode splitting compared to the inner $RW$. Typically, we target a $\approx 1.2$ GHz mode splitting to achieve phase matching. Indeed, the PhC affects the dispersion of the pump mode, splitting it in two while the dispersion of the remaining modes remains normal; see Fig. \ref{Fig2}e.

\begin{figure*}[ht]
\centering
    \includegraphics[scale=0.34]{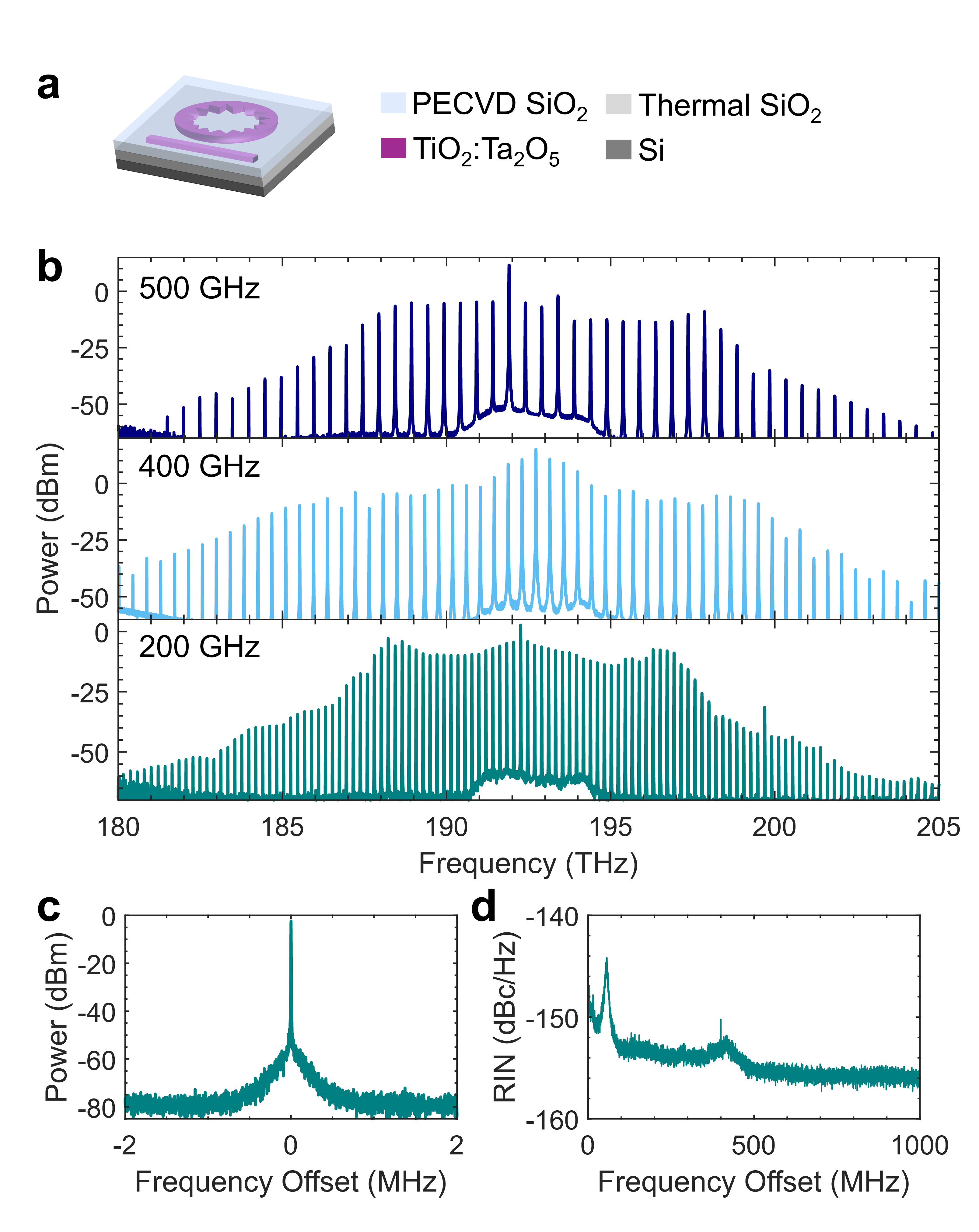}
    \caption{Soliton microcombs generated from SiO$_2$-clad titania-tantala PhCRs. (a) Illustration of the PhCR and film layer stack, with titania-tantala abbreviated as (TiO$_2$:Ta$_2$O$_5$). (b) Microcombs with a range of repetition rates achieved using different PhCR geometries. (c) Electrical spectrum of repetition rate noise and (d) relative intensity noise (RIN) of a 200 GHz repetition rate microcomb. }
    \label{Fig3a}
\end{figure*}

We test our ability to tailor microcomb spectra in SiO$_2$-clad PhCRs; see Fig. \ref{Fig3a}. We perform a similar experiment as we did in air clad PhCRs, where we systematically vary the geometry across devices to explore control over the repetition rate in normal dispersion microresonators. By adjusting the $RR$ from 42.6 \textmu m to 100 \textmu m, we demonstrate control over the repetition rate from 500 GHz down to 200 GHz; see Fig. \ref{Fig3a}b. Here, we present backward-propagating spectra, but we note that forward-propagating spectra are sometimes observed at larger detuning. Indeed, the soliton microcomb is stable, based on characterization of the repetition rate noise and relative intensity noise; see Fig. \ref{Fig3a}c and d, respectively. Using titania-tantala enables this systematic study because of its improved properties over pure tantala. In pure tantala with SiO$_2$ cladding, not only do the oxygen vacancy defects degrade $Q_i$, but they can contribute to unwanted photorefractive effects. Photorefraction presents as a transient change to the PhC mode structure that occurs when pumping a resonator, due to charge accumulation in defects \cite{Liu:2021_PhysRevLett}. These changes, which are time- and power-dependent, can interfere with phase matching and prevent microcomb generation. In titania-tantala, this effect is suppressed significantly due to its low defect density \cite{Carollo:2026}, allowing us to more systematically and reproducibly investigate soliton dynamics in SiO$_2$ clad PhCRs.

\begin{figure*}[ht]
\centering
    \includegraphics[scale=0.34]{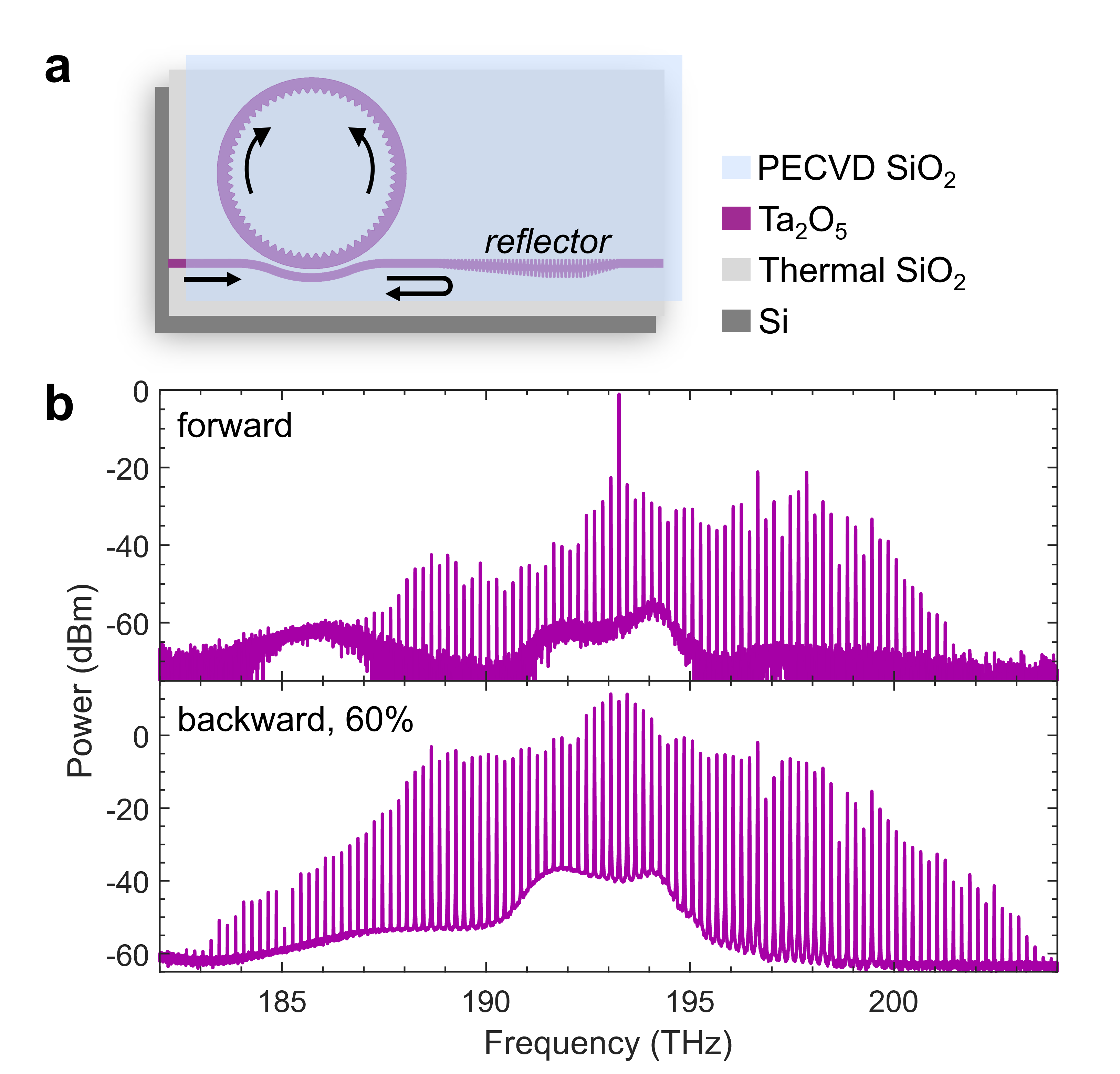}
    \caption{High-efficiency soliton microcombs generated from an SiO$_2$-clad, pure tantala PhCR with an integrated nanophotonic reflector. (a) Illustration of a tantala (Ta$_2$O$_5$) PhCR and a nanophotonic reflector, with arrows indicating the pump propagation directions. (b) Microcomb spectra in the forward (top) and backward (bottom) directions from pumping a PhCR with reflector.}
    \label{Fig3b}
\end{figure*}

A compelling advantage of the normal dispersion PhCRs that we study in this work is their potential to support high pump-to-comb conversion efficiencies, especially when implementing a nanophotonic reflector. With a reflector, we can increase the conversion efficiency in PhCRs beyond the theoretical limit of 50 \% \cite{Zang2025}. In Fig. \ref{Fig3b}a, we illustrate a PhCR with a nanophotonic reflector on the coupling waveguide, where we introduce a periodic modulation of the waveguide width that is designed to recycle a portion of the unused pump light back into the ring \cite{Zang2025}. This facilitates microcomb generation with an efficiency of 60 \%; see Fig. \ref{Fig3b}b. Our PhCRs tend to generate soliton microcombs primarily in the backwards direction, which is a consequence of the forward and backward coupling induced by the PhC. Here, the device is fabricated from pure tantala as part of our early efforts to develop the SiO$_2$ cladding process. Using titania-tantala instead could be a route to systematically studying high-efficiency combs with nanoreflectors and SiO$_2$ cladding in the future.

\begin{figure*}[!t]
\centering
    \includegraphics[scale=0.34]{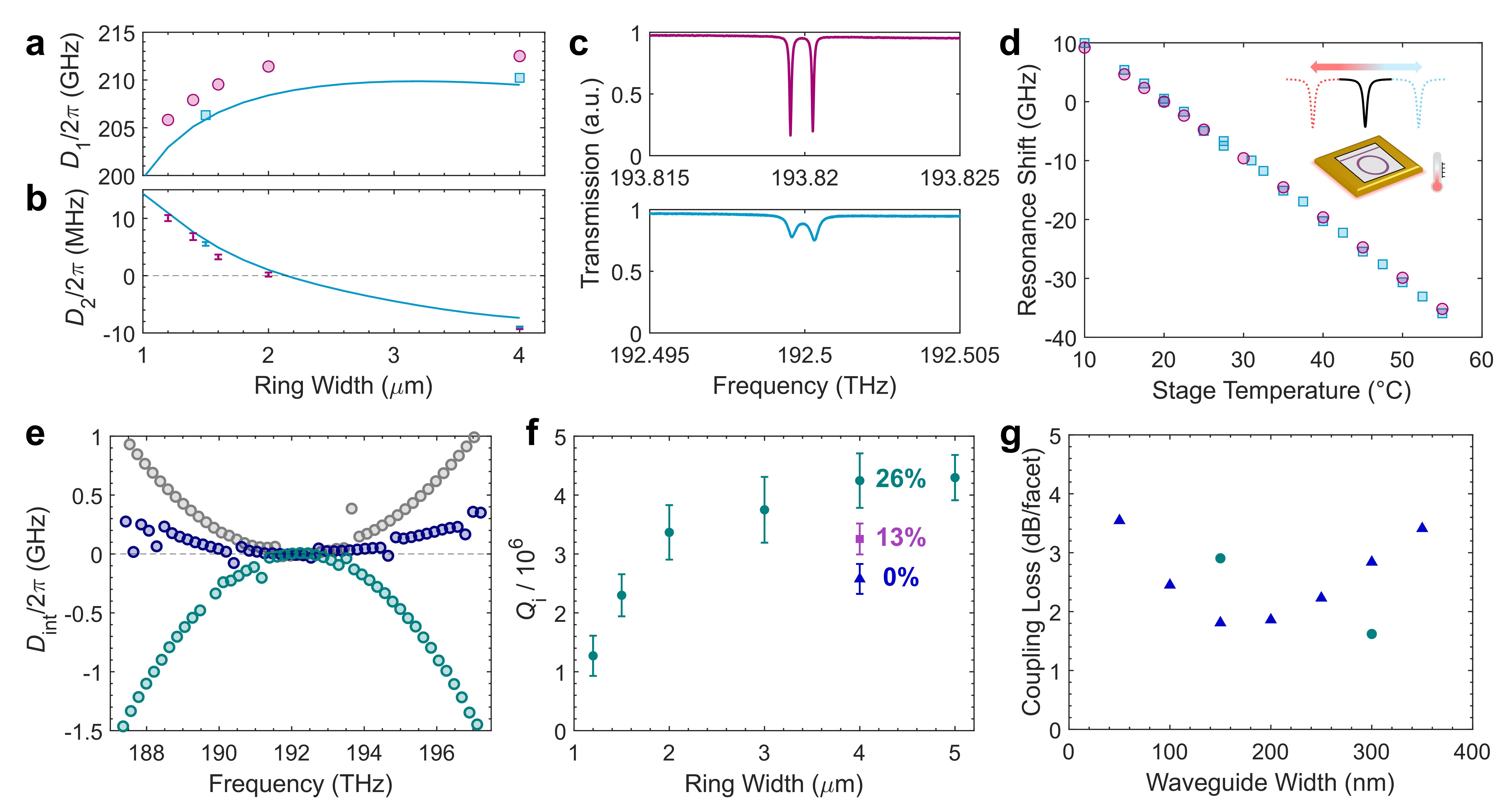}
    \caption{Dispersion engineering for tailoring microcomb spectral properties. (a) $D_1$ and (b) $D_2$ as a function of $RW$ for air clad titania-tantala microresonators. The solid lines are simulations, compared to measured values (blue points). Points in red show the effect of post-fabrication annealing on the dispersion. The dashed line in (b) marks $D_2/2\pi = 0$, and the error bars show the 95 \% confidence interval of $D_2$ obtained from the fits. (c) Effect of annealing on the PhC mode frequency and lineshape. The transmission trace before annealing is shown in blue, and after annealing is shown in red. (d) Characterization of the resonance frequency shift as a function of chip stage temperature for un-annealed (blue squares) and annealed (red circles) titania-tantala microresonators. Inset: schematic of the measurement. (e) $D_{int}$ for an SiO$_2$-clad titania-tantala microresonator with $RW = 1.5$ \textmu m (gray), $RW = 2$ \textmu m (blue), and $RW = 5$ \textmu m (green). The dashed line indicates $D_{\rm{int}} = 0$. (f) $Q_i$ as a function of $RW$ for SiO$_2$ clad tantala-based microresonators with various concentrations of TiO\textsubscript{2}: 0 \% (pure tantala, blue triangles), 13 \% (magenta squares), and 26 \% (green circles). We plot average $Q_i$ across several resonance modes, and we plot the standard deviation as error bars. (g) Fiber to chip coupling loss as a function of input waveguide width for titania-tantala (green circles) or pure tantala (blue triangles). }
    \label{Fig4}
\end{figure*}

So far in this work, we have validated that the tantala platform does support fully SiO$_2$-clad PhCRs for controlling soliton dynamics, particularly when the composition of the tantala film is modified to incorporate titania. Here, we investigate more properties of the titania-tantala films. To isolate the film properties, we pivot our focus back to air clad devices, but this time with titania-tantala. We begin by describing our dispersion engineering process. To achieve phase matching for microcomb generation, the cold-cavity resonance modes of a microresonator are designed to be non-uniformly spaced. We approximate the resonance mode frequencies to second order, as $\omega_\mu = \omega_0 + D_1\mu + (D_2/2)\mu^2$, where $\mu$ is the mode number relative to the pump mode, $D_1/2\pi$ is the free spectral range, and $D_2/2\pi$ is related to the GVD. We characterize how the mode frequencies deviate from an evenly-spaced grid with the integrated dispersion, $D_{int}(\mu) = \omega_0 - \omega_\mu - D_1\mu$. We characterize the mode structure by measuring transmission across the 1520 nm to 1600 nm wavelength range. We use the resonance mode frequencies to extract $D_1$ and $D_2$ for various various $RW$; see Fig. \ref{Fig4}a,b. Our results agree well with simulations, where we use the titania-tantala index of refraction in our finite-element analysis procedure for microresonator GVD. For the simulations, we use a Sellmeier fit of the index of refraction data (solid trace from Fig. \ref{Fig1}c) in the form,
\begin{equation}
    n^2(\lambda) = A + \frac{B_1\lambda^2}{\lambda^2-C_1^2} + \frac{B_2\lambda^2}{\lambda^2-C_2^2} + \frac{B_3\lambda^2}{\lambda^2-C_3^2}
    \label{eq:n}
\end{equation}
with wavelength, $\lambda$, in micrometers and coefficients $A = 2.371$, $B_1 = 0.031$, $B_2 = 2.180$, $B_3 = 8.958$, $C_1 = 0.342$ \textmu m, $C_2 = 0.223$ \textmu m, and $C_3 = 20.762$ \textmu m. To achieve a free spectral range near 200 GHz, we design the microresonator to have a $RR=100$ \textmu m and thickness of 570 nm. For a fixed $RR$ and thickness, adjusting the $RW$ controls $D_2$ to achieve either anomalous ($D_2>0$) or normal ($D_2<0$) dispersion. Adjusting the $RW$ introduces minor changes to $D_1$. We repeat this process for various $RR$ to target a range of repetition frequencies. Since annealing is optional in air-clad titania-tantala devices, except in applications requiring ultra-high $Q_i$, we investigate its effect on the dispersion. By comparing annealed (red points) to un-annealed (blue points) in Fig. \ref{Fig4}a and b, annealing increases $D_1$, while leaving $D_2$ unchanged. This is consistent with a reduction of the index of refraction upon annealing. This also causes a shift of the PhCR mode towards shorter wavelengths by $\approx 10$ nm, or $\approx 1.3$ THz; see Fig. \ref{Fig4}c. In PhCRs, we design the modulation period, $\Lambda = \pi RR /m$, to induce a photonic bandgap at a specific resonance mode with mode number $m=2\pi RR n_{\rm{eff}}/\lambda$, where $n_{\rm{eff}}$ is the effective index of refraction and $\lambda$ is the wavelength. Thus, a change in the index of refraction alters the wavelength of the split mode, which is important to consider when designing PhCRs. The transmission lineshape changes because of the increase in $Q_i$ with annealing. While we consider annealing to induce irreversible changes in the resonance frequencies, we achieve reversible tunability with the temperature of the chip; see Fig. \ref{Fig4}d. We demonstrate broad tuning of the resonance frequency as a function of stage temperature, and we estimate the thermo-optic coefficient of titania-tantala to be $1 \times 10^{-5}$ K\textsuperscript{-1}.

With our understanding of the factors that affect dispersion engineering in air clad titania-tantala, we use this process to further explore the design space with SiO$_2$ cladding. We present dispersion engineering capabilities in SiO$_2$-clad titania-tantala microresonators in Fig. \ref{Fig4}e. We demonstrate control over $D_{int}$, achieving either anomalous or normal dispersion depending on $RW$, for a fixed $RR$ = 100 \textmu m and thickness of 800 nm. Here, we achieve anomalous dispersion with $RW = 1.5$ \textmu m and $2$ \textmu m, and normal dispersion with $RW = 5$ \textmu m. Because $RW$ is an important design parameter, we systematically measure $Q_i$ as a function of $RW$ to understand how microresonator geometry affects optical loss; see Fig. \ref{Fig4}f. As we reduce the $RW$, the $Q_i$ gradually drops. This is likely caused by an increase in scattering, since the mode profile of the propagating light interacts more with the rough sidewall surfaces at small $RW$. This work focuses on normal dispersion resonators with large $RW$, so our PhCRs are not significantly impacted by scattering. Instead, our microresonator $Q_i$ is primarily limited by material absorption. We explore tantala-based films with different concentrations of titania to understand the effect that material composition has on microresonator $Q_i$ with SiO$_2$ cladding. Indeed, the titania-tantala films we use in the majority of this work, where the TiO\textsubscript{2} content is 26 \%, offer the highest $Q_i$ compared to pure tantala (i.e., 0 \% TiO\textsubscript{2}) and a tantala film with 13 \% TiO\textsubscript{2}. We anticipate 26 \% titania-tantala to have the lowest oxygen-vacancy defect density among the three films \cite{Fazio:20, Amato:23}, thus explaining its high $Q_i$. Besides high $Q_i$, tantala-based devices with SiO$_2$ cladding support low-loss fiber-to-chip edge couplers; see Fig. \ref{Fig4}g. To couple light onto the chip, we use lensed fibers to focus the light into an inversely-tapered waveguide. By sweeping the input waveguide width at the chip edge, we find that the optimal width for coupling loss below 2 dB/facet is near 300 nm in titania-tantala, or 150 nm in pure tantala, at the chip edge. This allows for efficient use of the pump power, supporting low laser power requirements.

\section{Conclusion}

In conclusion, we have demonstrated control over soliton microcomb dynamics through dispersion engineering in tantala-based nanophotonic microresonators. We fabricate PhCRs with ultralow-loss SiO$_2$ cladding, preserving nanoscale features necessary to enable high-efficiency microcombs with a wide range of repetition rates. Our low-temperature fabrication process ensures compatibility with a variety of photonic materials, while maintaining robust and stable performance, particularly when using titania-tantala films. These results pave the way for scalable, high-performance microcomb-based systems.

\section{Acknowledgments}
We acknowledge David Carlson from Octave Photonics for assistance with fabrication, and Jennifer Black from NIST for assistance with dispersion simulations. We thank Sarang Yeola and Shannon Duff for internal review. This research has been funded by the AFOSR FA9550-20-1-0004 Project Number 19RT1019, DARPA NaPSAC, DARPA QuICC (FA8750-23-C-0539), Northwest-AI-Hub, and NIST. Trade names provide information, not an endorsement. This is a work of the US government and is not subject to copyright in the US.

\section{Data Availability}
The data that support the findings of this study are available from the corresponding author upon reasonable request.

\printbibliography

\end{document}